# A Ligand-and-structure Dual-driven Deep Learning Method for the Discovery of Highly Potent GnRH1R Antagonist to treat Uterine Diseases

Song Li,[1][†] Song Ke,[4][†] Chenxing Yang,[4] Jun Chen,[4] Yi Xiong,[6] Lirong Zheng,[1] Hao Liu,[2,4,*] Liang Hong[1,2,3,4,5,6][*]

[1]School of Physics and Astronomy, Shanghai Jiao Tong University, Shanghai 200240, China

[2]Institute of Natural Sciences, Shanghai Jiao Tong University, Shanghai 200240, China

[3]School of Pharmacy, Shanghai Jiao Tong University, Shanghai 200240, China

[4]Shanghai Matwings Technology Co., Ltd., Shanghai, 200240, China

[5]Shanghai Artificial Intelligence Laboratory, Shanghai 200232, China

[6]School of Life Sciences and Biotechnology, Shanghai Jiao Tong University, Shanghai 200240, China

[†]These authors contributed equally: Song Li, Song Ke

[*]These authors jointly supervised this work: Hao Liu (chaohao2010@sjtu.edu.cn); Liang Hong (hongl3liang@sjtu.edu.cn)

## Abstract

Gonadotrophin-releasing hormone receptor (GnRH1R) is a promising therapeutic target for the treatment of uterine diseases. To date, several GnRH1R antagonists are available in clinical investigation without satisfying multiple property constraints. To fill this gap, we aim to develop a deep learning-based framework to facilitate the effective and efficient discovery of a new orally active small-molecule drug targeting GnRH1R with desirable properties. In the present work, a ligand-and-structure combined model, namely LS-MolGen, was firstly proposed for molecular generation by fully utilizing the information on the known active compounds and the structure of the target protein, which was demonstrated by its superior performance than ligand- or structure-based methods separately. Then, a in *silico* screening including activity prediction, ADMET evaluation, molecular docking and FEP calculation was conducted, where ~30,000 generated novel molecules were narrowed down to 8 for experimental synthesis and validation. In *vitro* and in *vivo* experiments showed that three of them exhibited potent inhibition activities (compound 5 IC50 = 0.856 nM, compound 6 IC50 = 0.901 nM, compound 7 IC50 = 2.54 nM) against GnRH1R, and compound 5 performed well in fundamental PK properties, such as half-life, oral bioavailability, and PPB, etc. We believed that the proposed ligand-and-structure combined molecular generative model and the whole computer-aided workflow can potentially be extended to similar tasks for *de novo* drug design or lead optimization.

# Introduction

Uterine fibroids, endometriosis, adenomyosis, cervical and uterine endometrial cancers, and ovarian cancer are all common proliferative diseases arising from gynecologic organs, especially uterine, that can harm the reproductive status of women [1, 2]. In elder women, dysmenorrhea is recurrent due to underlying issues such as uterine fibroids, adenomyosis, or endometriosis [3]. And endometriosis affects approximately 10% of all women during reproductive age, making it the second most common gynecological condition, after uterine fibroids [4]. In worldwide, more than 950,000 women per year are affected by cervical and uterine endometrial cancers, and more than 400,000 will die [2, 5]. Gonadotrophin-releasing hormone receptor (GnRH1R) is a promising therapeutic target to modulate the reproductive endocrine axis for the treatment of a variety of reproductive disorders, including uterine fibroids, endometriosis, and breast cancers [6, 7]. Up to now, several non-peptide GnRH1R antagonists have been well investigated clinically [8], including sufugolix [9], elagolix [10], relugolix [11] and linzagolix [12]. However, elagolix, which was the most clinical advanced nonpeptidic GnRH1R compound studied to date, appears to exhibit low oral bioavailability in both rats and monkeys (5.8% and 11%, respectively) [13], and a short half-life ($t_{1/2}$) that requires to be taken several times a day [8]. Pharmacokinetic (PK) and pharmacodynamic (PD) evaluations of relugolix revealed metabolic stability issues, and its Phase III clinical indication narrows to prostate cancer [14]. Linzagolix is a new investigational GnRH1R antagonist currently in clinical development with an appropriate half-life and high oral bioavailability [12], but the plasma protein binding (PPB) in human was estimated to be >99.4%, such high PPB value may arise drug-drug interaction issue if co-administered with another plasma protein binding agents simultaneously. Thus, the development of a new orally active small-molecule drug with a long half-life, high oral bioavailability, good PK/PD properties, and lower plasma protein binding is highly desirable.

Deep learning has recently been applied successfully in many fields, including drug discovery [15-17]. Breakthroughs were made in computer-aided drug design by artificial intelligence (AI), also known as AI-aided drug design (AIDD). An impressive deep generative model named GENTRL [18] was used to develop potent inhibitors against the discoidin domain receptor 1 kinase in 46 days, including 21 days to generate molecules for drug candidates. Generally speaking, the deep-generative approaches for designing drug molecules against a target protein can be classified into ligand-based and structure-based approaches. The ligand-based approach [18-26] requires the chemical and structural information of ligands, whose bioactivity has been proved in cellular, animal or even clinical experiments as reference to generate new molecules with optimal properties as shown in Fig. 1a. The structure-based methods [27-31] tend to generate molecules by optimizing the physical interaction of ligands with the structure of the target protein as Fig. 1b depicted. However, the ligand-based method could be biased towards the reference ligands in the training set and the generated molecules resemble them with low structural diversity (collapse to the yellow region in illustration of Fig. 1a), while the structure-based method may be hard to efficiently converge to a molecule with sufficiently high bio activity (explore to the blue region in illustration of Fig. 1b), particularly when examined by cellular or animal experiments. The chemical space of molecules generated through ligand-and-structure-based method was supported to be contained in the dotted circle in the illustration of Fig. 1c, which both cover the chemical space of bioactive ligands (yellow region) and explored ones (blue region). In this study, to effectively design a novel small molecule antagonist of the GnRH1R, we proposed a combined ligand-and-structure-based molecular generative model (LS-MolGen), as shown in Fig 1(d), by making use of a set of known bioactive ligands and the structure of GnRH receptor. Briefly, LS-MolGen is pretrained using the druglike molecules from ChEMBL with a conditional recurrent neural network (cRNN) and then pretrained knowledge is transferred via transfer learning (TL) to learn the known target-specific ligands with the same network. Finally, a fine-tune guided by reinforcement learning (RL) with docking score to obtain molecules of high binding affinity with target protein. Using LS-MolGen, we rapidly generated a novel molecule library of ~30,000 compounds and filtered it using proper physicochemical and medicinal metrics to get 10,000 molecules, which were further narrowed to 1,000 compounds by evaluating their bioactivity, and 200 compounds by evaluating the absorption-distribution-metabolism-excretion-toxicity (ADMET) via ADMETlab tool [32]. With comprehensive consideration of expert evaluation and detailed investigation of ligand-protein interactions by molecular docking analysis, 30 compounds were kept to perform free energy perturbation (FEP) calculation, and 8 compounds were finally selected for synthesis and experimental testing. In *vitro* and in *vivo* experiments showed that three of them had nanomolar (nM) activity (0.856 nM, 0.901 nM, 2.540 nM), and compound **5** performed well in fundamental PK properties, such as half-life, oral bioavailability, and PPB, etc. This work established an effective and efficient drug discovery pipeline by developing a ligand-and-structure dual-driven molecular generative model, LS-MolGen, making use of known information on bioactive compounds and structure of target protein.

## Results and discussion

**Architecture of LS-MolGen model.**

The architecture of LS-MolGen model is illustrated in Fig. 1d, which consists of three essential sub-models. The first model is an initial-stage model for pretraining of the general knowledge called the prior model (Fig. 1e), based on conditional RNN. The second model is a transfer model (Fig. 1f), based on the same architecture as the prior model, for transfer learning of the general knowledge to the focused knowledge via sharing the prior networks and reweighting the layers. The last one is an agent model (Fig. 1g) fine-tuned by reinforcement learning guided by molecular docking for the structure-based rational exploration of chemical space. Here, the docking score is obtained by LeDock [33], which is a direct measurement of the interaction strength between the ligand and the target protein by making use of the chemical and structural information on both the ligand molecule and the protein, particular its binding pocket. Hence, LS-MolGen is a ligand-and-structure dual-driven model which makes full use of the knowledge of both known activity ligands and protein structure to generate highly potent small molecules. Specially, to cope with the challenging multi-constrained optimization requirement, we imposed the constraint in the prior model by introducing a conditional token [34] (is or not GnRH1R antagonist) with the simplified molecular-input line-entry system (SMILES) string into the model training. It is also beneficial for the model to pay attention to the chemical space where the bioactivity of molecules satisfied a given threshold (pIC50 > 6) during the training process. Transfer learning is achieved by directly transferring the parameters of the prior model [25], which is computational cost effective. Finally, we introduced molecular docking score as a component of scoring function into reinforcement learning to guide the agent model exploring novel sub-areas in the chemical space, for example, enhancing the ability of the agent for scaffold-hopping that cannot be fully captured by the focused transfer model. And a scaffold similarity diversity filter (DF) [35] was used in the RL loop to enforce exploration. In the present work, LS-MolGen was trained to design potent antagonist against the target protein GnRH1R, the reference bioactive compounds include both the public bioactivity data retrieved from ChEMBL database and the private data provided by GeneScience Pharmaceuticals Co., Ltd., and the docking score in the RL loop was calculated using LeDock program (see *Methods and Materials* section for details).

**The performances of LS-MolGen model.**

We assembled three evaluation metrics: positive rate, internal diversity, and novelty, to comprehensively assess the performance of the molecular generative model. Here, the positive rate assesses the ability to generate potent high affinity (docking score < -8 kcal/mol) compounds, the internal diversity characterizes the structural diversity among generated molecules, while novelty quantifies the difference between the generated molecule as respect to the reference bioactive compounds. The detailed calculation methods of these three metrics are listed in *Methods and Materials*. To quantitatively compare LS-MolGen model with the ligand-alone model and the structure-alone model, we built a ligand-based model, QBMG, developed in Ref. [25] and a structure-based model modified from REINVENT [20] (denoted as REINVENT-m). The prior model was pretrained on the same dataset from ChEMBL for all these three methods, while QBMG shares the same transfer model as LS-MolGen and REINVENT-m utilizes the docking score as the scoring function to guide RL loop the same as in LS-MolGen.

For fair comparison of their performance, we trained and sampled 5,000 molecules from each model, and applied the evaluation metrics mentioned above for a comprehensive benchmark as shown in Fig. 2(j, k, and l). As evident by Fig. 2(a-l), the ligand-based model (QBMG) is capable to generate positive compounds with high similarity to the reference bioactive ligands, but at the same time, the generated molecules lack diversity among themselves and also lack the novelty as compared to the reference ligands. Structure-based model (REINVENT-m) could generate structurally and chemically very different and more positive compounds as compared to the QBMG. However, most of the generated molecules are quietly different from the known bioactive ligands and one can infer that the valuable experimental knowledge of those reference ligands was underutilized therein. Compared with QBMG and REINVENT-m, LS-MolGen has the highest positive rate and internal diversity with intermediate novelty. The intermediate novelty is an indirect evidence that LS-MolGen makes use of the information on both the chemical structure of the reference ligands and the structure of the target protein. Moreover, Fig. 2m visualizes the chemical space of molecules generated by distinct models using t-SNE [36], one can see that the experimentally-proved bioactive ligands overlap more with molecules from QBMG and LS-MolGen but are separated from the molecules generated by REINVENT-m. Importantly, QBMG covers the known chemical space of reference ligands well but lacks the capability to search for external chemical space, while REINVENT-m explores randomly without known knowledge as a guide, likely causing the molecule generated stuck in some local optimum of chemical space whose bioactivity does not meet the requirement. The model we developed, LS-MolGen exhibits a satisfactory result, that is, it can generate molecules that can both covers known chemical spaces of the reference ligands and explores new spaces under the guidance of the structure of the target protein and its interaction with the ligands. In summary, LS-MolGen outperforms the other two methods and is more suitable for molecular generation in drug design.

**Binding mode between the target protein and ligand was achieved by induced-fit docking.**

The target protein of the present work is the GnRH receptor, a seven-transmembrane domain G-protein coupled receptor (GPCR) belonging to the class A family of GPCRs, playing an important role in the hypothalamic-pituitary-gonadal reproductive endocrine axis. Recently, the crystal structure of human GnRH1R in complex with the antagonist drug elagolix was reported in Ref. [37]. As revealed in it, the overall binding pocket in GnRH1R is defined by TM2, TM3, TM5, TM6, TM7, and the N-terminus, forming a highly hydrophobic binding site, as shown in Fig. 3a. Particularly, the N-terminus of GnRH1R potentially participated in ligand binding as its N27 interacted with ligand elagolix. The pyrimidine ring of elagolix forms a hydrophobic $\pi-\pi$ interaction with the aromatic side chain of Y283$^{6.51}$, and the oxygen atom on the pyrimidine ring is engaged in binding with the side chains of residues K121$^{3.32}$ and D98$^{2.61}$, forming a polar interaction network in the elagolix binding site, which contributes to high-affinity ligand binding. Such binding mode of GnRH1R-elagolix was further confirmed by molecular modeling through clinically approved drug relugolix and its analog sufugolix. Analogously, in this work, we carried out molecular docking of linzagolix, which is also a small molecule antagonist but lacks the ligand-protein co-crystal structure. By docking linzagolix into the binding pocket of GnRH1R-elagolix [37] as illustrated in Fig. 3b, a surprisingly low docking score (-9.316 kcal/mol) was observed, much lower than the redocking score (-14.027 kcal/mol) of elagolix. Further induced-fit docking (IFD) revealed a novel and strong ligand-binding mode of linzagolix as shown in Fig. 3c, yielding a much higher docking score (-11.397 kcal/mol). More interesting, comparing the poses of linzagolix obtained by docking at the binding pocket of elagolix and by the induced-fit docking, we found that the thienopyrimidine ring of ligand undergoes a large conformational flip, from the outside which is close to the N-terminus (Fig. 3b) to the inside which approaches the residue of Y283$^{6.51}$ (Fig. 3c). As a result, the thienopyrimidine ring instead of the benzene ring in the middle of linzagolix forms a hydrophobic $\pi-\pi$ interaction with Y283$^{6.51}$ and a hydrogen bond with K121$^{3.32}$, and the oxygen atoms on these two benzene rings of the ligand form three strong hydrogen bonds with the N27 residue. In addition, conformational rearrangement of the side chain of Y290$^{6.58}$ was observed in the IFD structure, whereas the aromatic side chain changes from being repelled outward (Fig. 3b) to turn towards the thienopyrimidine and forms a hydrogen bond with the oxygen atom on it (Fig. 3c). All the above interactions render a novel and strong binding mode between the target protein and ligand, which could be informative on the design of GnRH1R antagonist. This IFD derived binding mode of linzagolix was used as a target protein structure in the third sub-model of LS-MolGen (Fig. 1g) to generate molecules with high docking score.

**Discovery of GnRH antagonist molecules based on LS-MolGen**

In this work, we constructed an AI driven pipeline for the discovery of GnRH1R antagonist molecules based on the proposed molecular generative model. As shown in Fig. 4b, firstly, the LS-MolGen model was developed to generate a molecule library including about 30,000 novel compounds, and then filters were automatically applied to exclude compounds with inappropriate molecular weight, structural alerts, and false-positive group [38]. To evaluate the properties of these molecules, the physicochemical and medicinal properties, for example, molecular weight (MW), water-octanol partition coefficient (LogP), quantitative estimate of drug-likeness (QED), and synthetic accessibility (SA) score were calculated as shown in Fig. S1 in Supplementary Information (SI). And to analyze the chemical scaffolds, we counted the distribution of the Bemis-Murcko (BM) scaffold for these molecules. The 10,000 molecules contained 6196 BM scaffolds, some examples of which are shown in Fig. 4a. The drawn-out top 4 counts of scaffolds are novel scaffolds except scaffold **2**, mimicking linzagolix. This result also indicates that LS-MolGen can explore new chemical scaffolds while recovering the known positive scaffold.

Subsequently, in *silico* screening was performed to identify the potent antagonist molecules. To prioritize molecules for the binding affinity of GnRH1R, we built a Fully-connected-neutron-network-based model to predict activity (see *Methods and materials* and SI for details of the model). The top 1,000 ranking compounds were collected for further screening using ADMETlab website [32], and the compounds with predicted PPB lower than 95% and bioavailability larger than 30% as well as those without metabolic defects were selected. With comprehensive consideration of the evaluation by the medicinal chemist and of the molecular docking scores, we further selected 30 molecules to perform the calculation of the free energy perturbation (FEP) (see Fig. 4c or Table. S2). Finally, by evaluating of the docking scores and FEP results, we were able to rank these 30 compounds, the top 3 compounds (Compound **1**, **3** and **5**) of which were displayed in Fig. 4c. We found that compound **3** and compound **5** share the same scaffold as scaffold **2** (see Fig. 4a), and compound **1** has conducted a scaffold hopping from a benzene to a pyridine. More importantly, the docking poses and the interaction with surrounding key residues of these three molecules match the induced-fit docking pose of linzagolix with the protein as shown in Fig. 4c, which validate the novel binding mode revealed in the previous section, providing us the confidence to choose them as lead compounds for further synthesis and activity validation.

**Synthesis and activity validated in *vitro/vivo***

Supplemented with advice from chemist in structural optimization, we finally synthesized and activity tested eight compounds in *vitro*. As an example, the synthesis of the target compound **5** is depicted in Fig. 5 (for other compounds please see SI). In brief, the intermediate 5b was obtained by etherifying the starting materials 5a using 1-Bromo-2-methoxyethane, and then, its nitro group was reduced to amino group via zinc powder to give intermediate 5c. The subsequent intermediate 5d was obtained by a urea-forming reaction between the amino group of 5c with 4-((phenoxycarbonyl)amino)thiophene-2,3-dicarboxylate. Finally, the target compound **5** was obtained through self-cyclization and ester hydrolysis of intermediate 5d in lithium hydroxide solution.

Encouragingly, all these eight compounds strongly inhibited GnRH-stimulated calcium release from HEK 293 cells expressing human GnRH receptors in FLIPR assay (see Table. 1). Particularly, three of them (compounds **5**, **6** and **7**) had nanomolar bioactivity, and two of them shown high affinity with IC50 = 0.856 nM (compound **5**) and 0.901 nM (compound **6**). We noted that compounds **1**, **3**, and **5** are designed by LS-MolGen and directly screened out based on our workflow (Fig. 4b), and compounds **2**, **4**, and **6-8** are modified from compounds **1**, **3**, and **5**, respectively by the medicinal chemist. Based on the high potent activity assay (see Fig. 6a), compounds **5** and **6** were further selected for in *vitro* and in *vivo* pharmacokinetics (PK) test in rats. Remarkedly, in *vitro* enzyme activity test show that compound **5** has the best performance, which does not act as any inhibitors for any of human CYP1A2, CYP2C9, CYP2C19, CYP2D6 CYP3A4-M enzymes (Table. S2), and is stable in liver microsomal stability test across human, rat and dog with half-life larger than 145 min (Table. S4-6), and a favorable in *vivo* PK profile in rats after single oral dose of this compound at 3 mg/kg in Female SD Rat was also obtained as shown in Fig. 6b. Briefly, the mean $C_{max}$ = 6403 ng/ml, $AUC_{INF\_pred}$ = 111425 h*ng/ml, $t_{1/2}$ = 11.1 h, and bioavailability F = 48.7%. More detailed PK profiles of compound **5** and **6** were presented in Table. S9-10.

As mentioned before, linzagolix, as an acid compound may exhibit high plasma protein bound fraction, which could possibly impair the stability of plasma-free concentrations leading to adverse pharmacologic consequences. We highlight that, the compound **5** we designed here, showed a lower human PPB fraction as compared to linzagolix (96.81% v.s. >99.4%) (see Table. S8), which may reduce the fluctuation of plasma-free drug concentration and help stabilize drug efficacy. Thus, compound **5** was demonstrated to be the highly potent antagonist molecule with high oral bioavailability, good PK properties, and lower PPB. Of course, determining whether compound **5** can be eventually advanced to clinical trials needs further experimental study.

# Conclusion

In the present study, we have developed LS-MolGen, a novel molecular generative model for *de novo* drug design that generates molecules bases on the knowledge from both bioactive ligands and structure of target protein. LS-MolGen consists of three essential sub-modules, which are conditional RNN, transfer learning and reinforcement learning, and exhibits superior performance in generating new molecules with high docking score, large internal diversity, and great novelty with respect to the reference compounds.

We constructed an AI driven pipeline for developing potentially promising drug candidates for GnRH1R. ~30,000 molecules were first generated by LS-MolGen, and then virtually screened by activity and ADMET evaluation to be reduced to ~100 potent compounds. Further molecular docking, free energy perturbation and rational structural optimization helped us select eight compounds for synthesizing and activity testing. Remarkably, three of them exhibited high potent inhibitory activity (compound **5** IC50 = 0.856 nM, compound **6** IC50 = 0.901 nM, compound **7** IC50 = 2.54 nM) against GnRH receptor. And in *intro* and in *vivo* PK test in rats show that compound **5** did not inhibit enzymes of the human CYP family, performed well in fundamental PK properties, such as half-life, oral bioavailability and presented superior PPB etc., which could be a good candidate for further drug development.

To summarize, this study developed ligand-and-structure dual-driven deep learning molecular generative method for de novo generation of molecules. Using this generation model, we took the GnRH1R as an example to go through a complete computer-based workflow for in *silico* screening including activity prediction, ADMET prediction, molecular docking and FEP calculation, to design potent antagonist molecule. This workflow starts from 30,000 novel compounds generated and then narrows the selection down to less than ~10 for further synthesis and experimental testing, finally yielding a promising lead compound. The detailed features of the generation model (LS-MolGen) and the workflow developed here can be tailored in the future to carry out many different assignments on *de novo* drug design or lead optimization.

## Methods and materials

### Datasets

The dataset for pretraining in the prior model was obtained from the work of Olivecrona et. al.[20], including 1.5 million structures selected from ChEMBL [39] where the molecules were restrained to contain between 10 and 50 heavy atoms and elements $\in$ {C, H, O, N, P, S, F, Cl, Br, I}.

The bioactivity dataset for GnRH1R includes the public bioactivity data retrieved from ChEMBL database and proprietary data provided by GeneScience Pharmaceuticals Co., Ltd. The data was standardized, yielding 1010 compounds with unique bioactivity values and then used in construction of activity prediction model. The compounds with high activities were used for the transfer learning of LS-MolGen. The details can be found in SI.

The crystal structure of human GnRH1R in complex with antagonist elagolix was retrieved from PDB (ID: 7BR3) [37].

### Architecture of LS-MolGen

A brief overview of the entire architecture is illustrated in Fig. 1d. A general prior model was first trained to conditionally generate the desired molecules, then the transfer model conducts TL based on the knowledge transferred from prior, and the subsequent integration with RL.

***The prior model.*** The architecture of the prior model is provided in Fig. 1e. This model is expected to learn to generate molecules with desirable properties encoded by a set of conditional tokens (as a constraint code $c$). In this work, the conditional tokens are "is_GnRH" and "not_GnRH" to determine whether a molecule is or not a GnRH1R antagonist, judging by the predicted pIC50 (pIC50 > 6 for "is_GnRH" and pIC50 < 6 for "not_GnRH"). Conventionally, $pIC50 = 9 - \log(IC50[nM])$. The standard RNN model [20] was modified for molecular generation by adding a joint embedding of constraints and SMILES. Maximum likelihood estimation with a given $c$ was employed as followed to train the conditional RNN with 1024 Gated Recurrent Units (GRU) in

each of three layers:

$$NLL(s) = -\sum_{i=1}^{n} \log P(x_i \mid x_{<i}, c)$$

***The transfer model.*** This model requires a pretrained prior model with a generative capacity and the potential to sample compounds from a rather vast chemical space. And the prior is subjected to transfer learning with a smaller set of known bioactive ligands. As a result, it will produce compounds similar to the existed target ligands with a higher probability. The transfer learning network is the same as the prior network.

***The agent model.*** The agent model is an exploration model which shares identical architecture and vocabulary with the transfer. Essentially, the agent was initialized with the transfer network at the beginning of the RL. We adopted the RL algorithm in REINVENT2.0 [35] to fine-tune the transfer model, and built a customized scoring function for the molecular optimization:

$$Score(s) = \begin{cases} -\text{Docking Score}(s), & \text{unsatisfy DF} \\ 0, & \text{satisfy DF} \end{cases}$$

An DF is used to evaluate whether the SMILES string has been sampled before or whether it satisfies the DF policy. The Score will be set to 0 if the DF filters determine that the sampled compound already exists or if there are too many compounds of the same scaffold and their number exceeds our predefined threshold. Analogously, the $NLL(s)$ for the given string $s$ is also calculated by the agent:

$$NLL(s)_{\text{Augmented}} = NLL(s)_{\text{transfer}} - \sigma \cdot Score(s)$$

The Score is multiplied by $\sigma$ which is a scalar coefficient used for scaling up the scoring function output to the same order of magnitude as the $NLL$. And the loss is calculated as the squared difference between the likelihood of agent and the augmented likelihood:

$$loss = [NLL(s)_{\text{Augmented}} - NLL(s)_{\text{Agent}}]^2$$

The final component of the RL loop as shown in Fig. 1f is inception, which is a modified version of experience replay. The purpose of inception is to "memorize" previously well-scored compounds and to randomly present a subset of them to the agent. This serves as "reminder" of which SMILES were better and speeds up the learning process. We noted that the compound generation is done during the process of RL and not after, which is a notable difference between TL and RL as we do not use the end state of the model.

**Evaluation metrics for generative model**

***Positive rate***: the proportion of generated molecules ($G$) with high affinity (docking score <-8 kcal/mol) docked in GnRH1R. Here, the high-throughput molecular docking was performed by LeDock program [33], and the structure of receptor was identified from GnRH1R-elagolix complex (PDB ID: 7BR3).

$$\text{Positive rate} = \frac{1}{n}\sum_{g} 1[\text{docking score} < -8]$$

***Diversity:*** calculated based on the Tanimoto distance $\text{sim}(x, y)$ with respect to the Morgan fingerprints of a pair of sampled molecules, as the following equation:

$$\text{Diversity} = 1 - \frac{2}{n(n-1)}\sum_{x,y} \text{sim}(x, y)$$

***Novelty:*** calculated the similarity between the generated molecules ($G$) and reference molecules ($R$) in the training set as follows. For each molecule $g \in G$, the nearest neighbor molecule $r \in R$ is selected from the training set.

$$\text{Novelty} = \frac{1}{n}\sum_{g} 1[\text{sim}(g,r) < 0.4]$$

**The activity prediction model**

In bioactivity dataset, some of the compounds did not have the exact IC50 value but a range, such as "> 1000 nM", which is not suitable for the regression model. To fully use the information of these compounds, both classification and regression models were built to predict the affinity of the compound. The molecular fingerprints, such as CDK, ECFP4, MACCS, and PubChem, were applied for molecular representation. The metrics for both classification and regression tasks were area under ROC curve (AUC) and root mean square error (RMSE). The hyperparameters, such as batch size and learning rate, along with the different fingerprints were determined according to the model performance in five-fold cross-validation. A total of 50 epochs were run and the parameters of the models were saved for the epoch with the best performance. In addition, the ensemble method was used in this work, that is, the parameters of all five models in cross-validation were saved, and the final result was the averaged prediction values of the five models. Details can be found in SI.

**Molecular docking and induced-fit docking**

*Molecular docking*. The molecular docking for the scoring in RL and positive rate calculation in model performance section was conducted via LeDock program [33], and the molecular docking for the structure-activity analysis in this work was performed in the Maestro suite (https://www.schrodinger.com). PDB structure of 7BR3 was preprocessed and energy minimized using the Prep module. The binding site grid was generated around the elagolix binding site with similar in size to elagolix. Docking poses were generated by standard-precision (SP) Glide runs using the optimized ligand structure. The docking results are displayed by PyMOL software.

*Induced-fit docking.* The IFD was also performed in the Maestro suite with the Induced-fit Docking module. Following the standard protocol which generates up to 20 poses with force field of OPLS3e, and docking ligands similar in size to elagolix within the same GnRH1R-elagolix complex receptor. We noted that the linzagolix induced-fit structure of GnRH1R was used to generate molecules by LS-MolGen and the sequential molecular docking structure-activity analysis.

**FEP calculation**

The initial protein structure used for FEP calculation input was the linzagolix induced-fit structure of GnRH1R. The linzagolix was used as a reference molecule to determine the binding position of the ligands in the protein pocket, and other ligands were superimposed on the position of the reference molecule based on the maximum common substructure (MCS). The membrane in the simulation system was established in the CHARMM_GUI website, and the ratio of cholesterol and POPC on the membrane was 1:9. Each system was solvated in a cube box of TIP3P water model. The cross-sectional size of the membrane was 10nm*10nm, and the protein-ligand complex was placed in the center of the membrane. The net charge of the system was neutralized by adding Na+ and Cl- ions.

The FEP calculation was performed on the GROMACS software. All simulations used Amber14SB_IDLN force field for proteins, GAFF2 force field for ligands. The simulation flow of protein-ligand and membrane complexes is as follows. First, the system energy was minimized by 100,000 steps of steepest descent. The simulation flow of protein-ligand and membrane complexes is as follows. First, the system energy is minimized by 100,000 steps of steepest descent. Then under the ensemble of NVE and NVT, respectively, a molecular dynamics simulation of 20,000 steps was performed. After 400 ps equilibration at constant pressure, the production MD simulations were conducted in the NPT ensemble at 310 K. All simulations during FEP calculation used a leap-frog integrator with a 1 fs timestep for NVE and NVT ensemble, 2 fs for NPT and production. Cartesian restraints were applied to ligands and protein heavy atoms with a force constant 1000 kj mol$^{-1}$ nm$^{-2}$ for NVE ensemble. The long-range electrostatic interactions were calculated with Partial Mesh Ewald (PME) method. The SHAKE algorithm was used to constrain the lengths of bonds with hydrogen atoms.

In this work, 16 λ windows are set for each perturbation, and the simulation time for each window is set to 5 ns. The multistate Bennett acceptance ratio (MBAR) method was used for estimating expectations and free energy differences from equilibrium samples from multiple probability densities

**GnRH1R antagonist activity assay**

Compounds antagonist activity against GnRH1R was measured via FLIPR Calcium 6 Assay (Molecular Devices, R8191) and Flp In-293-GnRH Stable Pool cell lines were used for the IC50 test. Cells were cultured with the following medium (DMEM + 10% FBS +2mMGlutaMAX + 1x Penicillin-Streptomycin (PS) + 200μg/ml Hygromycin B). A 384-well plate was seeded with cells at a density of 12000 cells per well (25μl of cell seeding medium in 384-well cell culture plate (Corning, 3764), seeding medium: DMEM + 10% FBS +2mMGlutaMAX) and allowed to attach overnight. On the experiment day, the culture medium was carefully removed from the wells and 40 μL assay buffer (1 × Hank's balanced salt solution with 20 mM HEPES, pH 7.4) containing Calcium-6 dye was added in each well. Dilute GnRH1R with assay buffer to 180nM (6X), transfer 40μL to a 384-well plate (Corning, 3657). Reference and test compounds were dissolved in dimethyl sulfoxide (DMSO) and diluted with the assay buffer (no dye) at required concentrations. Take the 384-well cell culture plate out from incubator and place it at RT for 10min. Add 10μL of 6X compound working solution to the indicated wells and place the plate at RT for 30min, add 10μL of 6X GnRH by FLIPR Tetra to the 384-well cell culture plate and collect the data immediately (Molecular Devices, Sunnyvale, CA).

**CYP Inhibition Evaluations of Compound 5 in Human Liver Microsomes**

Water used in the assay and analysis was purified by ELGA Lab purification systems. Potassium phosphate buffer (PB, concentration of 100mM) and MgCl2 (concentration of 33mM) were used. Test compound (compound **5**) and standard inhibitors (α-naphthoflavone (CYP1A2), sulfaphenazole (CYP2C9), (+)-*N*-3-benzylnirvanol (CYP2C19), quinidine (CYP2D6), and Midazolam(CYP3A4)) working solutions (100×) were prepared. Microsomes were taken out of a freezer (–80 °C) to thaw on ice, and returned to the freezer immediately after use. 20μl of the substrate solutions was then added to corresponding wells, 20 μl PB was added to blank wells, and 2 μl of the test compounds and positive control working solution was added to corresponding wells. Next, we prepared a working solution of human liver microsomes (HLM), and 158 μl of the HLM working solution was added to all wells of the incubation plate. The plate was prewarmed for approximately 10minutes in a water bath at 37 °C. Then, reduced nicotinamide adenine dinucleotide phosphate (NADPH) cofactor solution was prepared and 20 μl NADPH cofactor was added to all incubation wells. The solution was mixed and incubated for 10minutes in a water bath at 37 °C. At this point, the reaction was terminated by adding 400 µl cold stop solution (200 ng/ml tolbutamide and 200 ng/ml labetalol in acetonitrile (ACN)). The samples were centrifuged at 4,000 r.p.m. for 20 minutes to precipitate protein. Then, 200 µl supernatant was transferred to 100 µl HPLC water and shaken for 10minutes. XL fit was used to plot the per cent of vehicle control versus the test compound concentrations, and for nonlinear regression analysis of the data. IC50 values were determined using three- or four-parameter logistic equation. IC50 values were reported as >50 µM when percent inhibition at the highest concentration (50 µM) was less than 50%.

**Pharmacokinetic studies.** The procedures that are applied on animals in this protocol were be approved by PharmaLegacy Laboratories IACUC. The pharmacokinetic profiling of compound **5** was performed on female SD rats (>200g). Then, we performed i.v. (1mg/kg) and p.o. (3mg/kg) administration of compound **5**. Each group consisted of three rats. The vehicle formulation of I.V. was 5%DMSO+15%Solutol HS 15+80% SBE-β-CD (20% in water, w/v), and the vehicle formulation of PO was 2%DMA/97.4% (0.5%CMC-Na in water)/0.6% 2N NaOH. All blood samples (approximately 110 µl blood per time point) were transferred into prechilled commercial K2-EDTA tubes, and then placed on wet ice. The blood samples were immediately processed for plasma by centrifugation at approximately 4 °C, at 450 rpm for 5 min and centrifuged at 4000 rpm for 10 min. 30 µL sample was supplemented with 150 µL IS (Propranolol, Tolbutamide, Glipizide, Osalmid, each 200 ng/mL in ACN). After shaked at 450 rpm for 5 min and centrifuged at 4000 rpm for 10 min, the supernatant 100 µL with 100 µL H2O was transfered into 96 plates until LC/MS/MS analysis. Plasma concentration versus time data was analyzed by non-compartmental approaches using the Phoenix WinNonlin® 8.2 (NCA) software program.

## Acknowledgments

This work was supported by the grants from the National Science Foundation of China [grant numbers 11504231, 21873101, 31630002, 32030063], the Innovation Program of Shanghai Municipal Education Commission [2019-01-07-00-02-E00076], and the student innovation center at SJTU. The authors acknowledge the Center for High Performance Computing at Shanghai Jiao Tong University for computing resources, and specially acknowledge Yue Huang, Guangyuan Shen and Xiaoyun Fu from GeneScience Pharmaceuticals for their tremendous supports in medicinal chemistry consultation as well as wet lab experiments coordination and data collection.

## Declaration interests

Song Ke, Chenxing Yang, and Jun Chen are employees of Shanghai Matwings Technology Co., Ltd., Shanghai. Other authors declare no competing interests.

## Data availability

The datasets of bioactive ligands and generated molecules in the current study are available upon reasonable request by contacting the corresponding authors.

## Author contributions

L. H. and H. L. designed and supervised the project. S. L. and S. K. contributed equally in this work. S. L. designed and implemented the deep learning model and performed the model training. S. K. organized the experimental results in *vitro* and in *vivo*. C. Y. and J. C. performed the calculation of FEP. H. L., S. L., and S. K. discussed and analyzed the data. S. L. wrote the manuscript. L.H., H. L., S. K. and Y. X. revised the manuscript.

**Table. 1.** *In vitro* antagonist activities of compounds **1-8** against GnRH1R

| Compound | Molecular structure | IC50 (nM) |
|---|---|---|
| 1 | 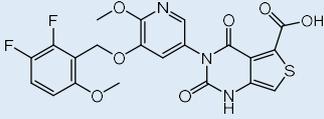 | 106.8 |
| 2 | 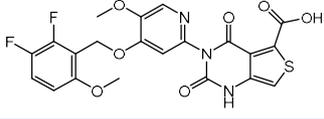 | 335.2 |
| 3 | 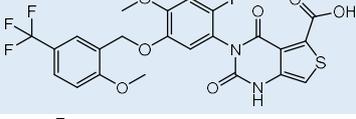 | 10.30 |
| 4 | 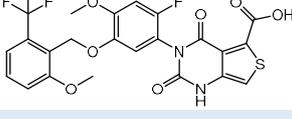 | 15.16 |
|  | 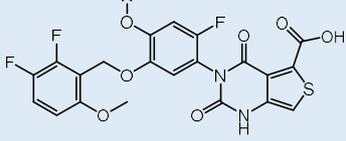 |  |
| 5 | 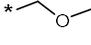 | 0.856 |
| 6 | 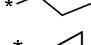 | 0.901 |
| 7 | 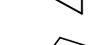 | 2.540 |
| 8 | 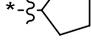 | 17.84 |

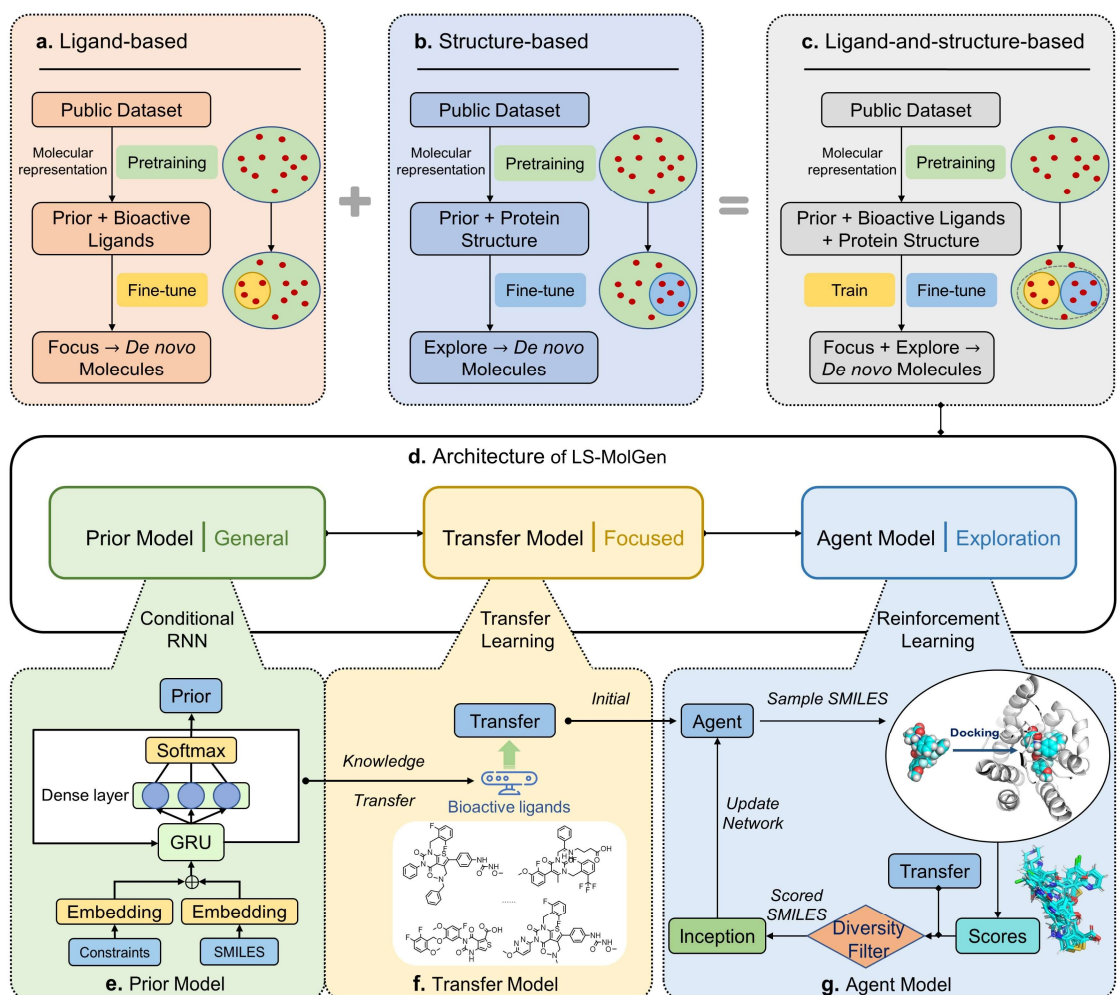

**Fig. 1.** The concept of ligand-based and structure-based methods, and the architecture of LS-MolGen approach. **a,** The concept of ligand-based method, **b,** structure-based method, and **c,** complementary ligand-and-structure-based method. Illustration in a, b, and c, the green region represents the available chemical space, the yellow region represents the chemical space of known bioactive ligands, the blue region represents the explored chemical space via RL, and the dotted circle contains the complementary chemical space of both known bioactive ligands and explored. **d**, The pipeline of the LS-MolGen approach. **e**, The architecture of the prior model with conditional RNN. **f**, The illustration of transfer learning model. **g**, Exploration of chemical space of reinforcement learning combine with molecular docking.

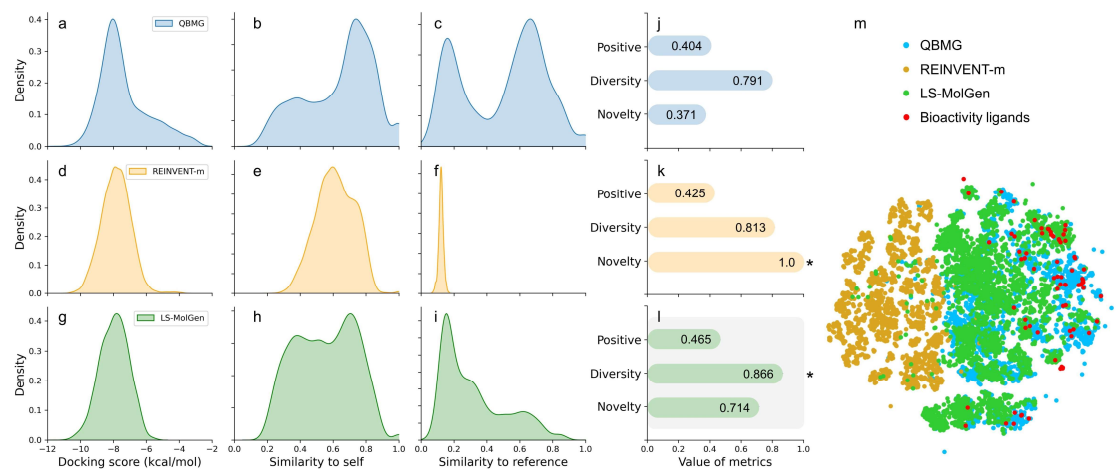

**Fig. 2.** The performance of distinct generative models. **a**, **b**, and **c**, is the distribution of docking score, similarity to themselves, and similarity to the reference compounds with known bioactivity respectively for molecules generated by QBMG. So are **d**, **e**, and **f**, for molecules generated by REINVENT-m, and **g**, **h**, and **i**, for molecules generated by LS-MolGen. **j**, **k**, and **l**, are the values of evaluation metrics for the molecules generated by QBMG, REINVENT-m, and LS-MolGen respectively. **m**, is the chemical space visualization using t-SNE dimensionality reduction.

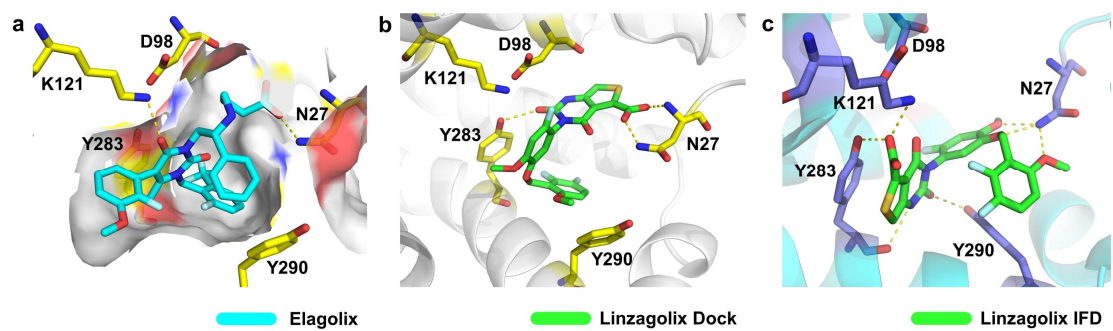

**Fig. 3.** Structural analysis of antagonist in GnRH1R reveals a novel and strong binding mode. **a**, is the x-ray crystal structure of GnRH1R bound to elagolix (PDB ID: 7BR3), the binding pocket is shown as surface and the key interacting residues are shown as sticks. **b**, and **c**, is the obtained binding mode of GnRH1R-linzagolix by docking and induced-fit docking linzagolix into the GnRH1 receptor, respectively. The dotted line represents hydrogen bond.

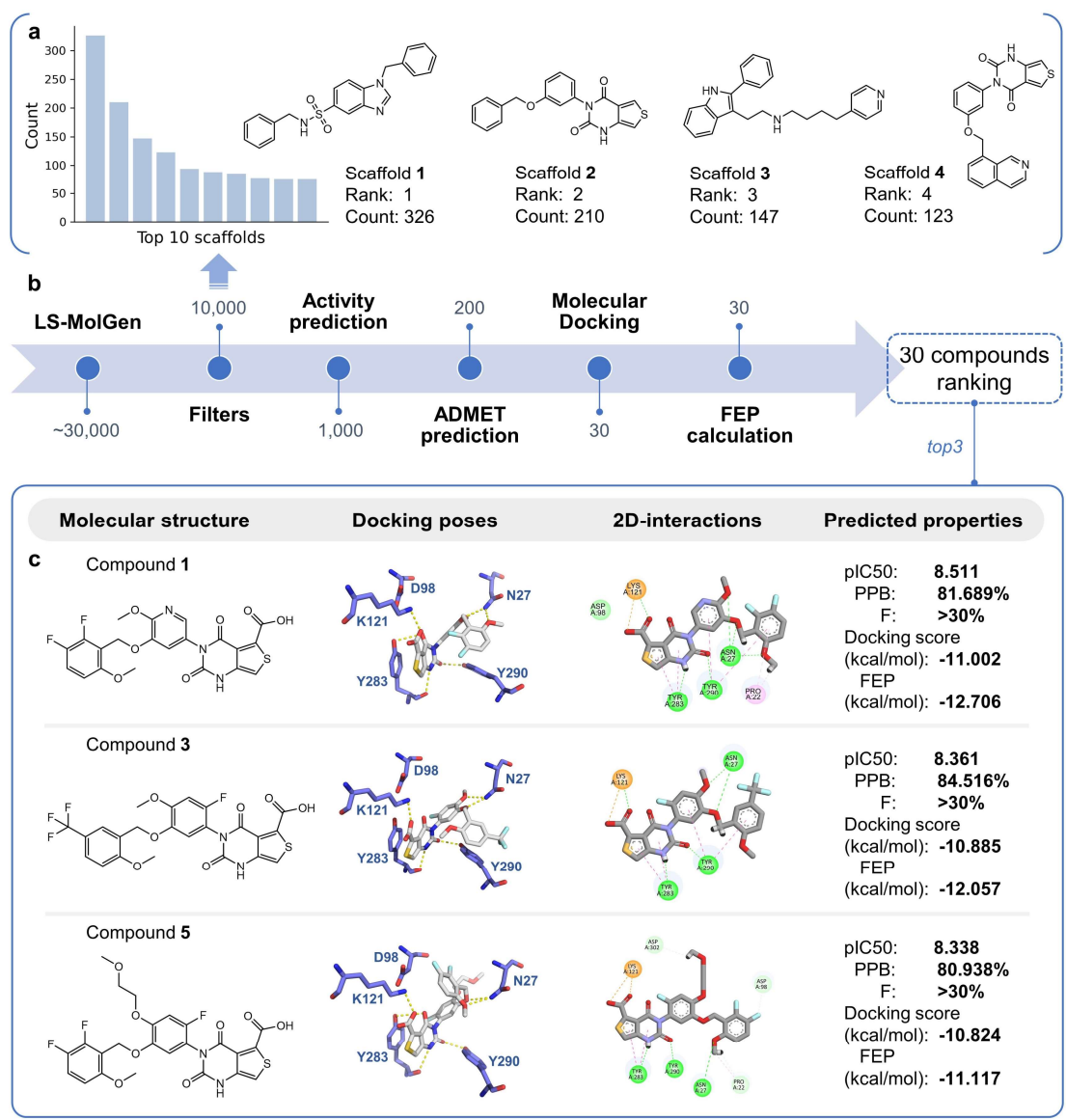

**Fig. 4.** The workflow of the discovery of GnRH antagonist molecules based on LS-MolGen. **a**, Count distribution of LS-MolGen generated scaffolds and examples of top scaffolds. **b**, The workflow for the screening of GnRH antagonist molecules. **c**, Analysis of top three promising compounds from molecular ranking. The first column: molecular structures of compounds. The second column: docking poses in the linzagolix induced fit structure of GnRH. The third column: two dimensional interactions between the molecules and the protein. The last column: the predicted molecular properties.

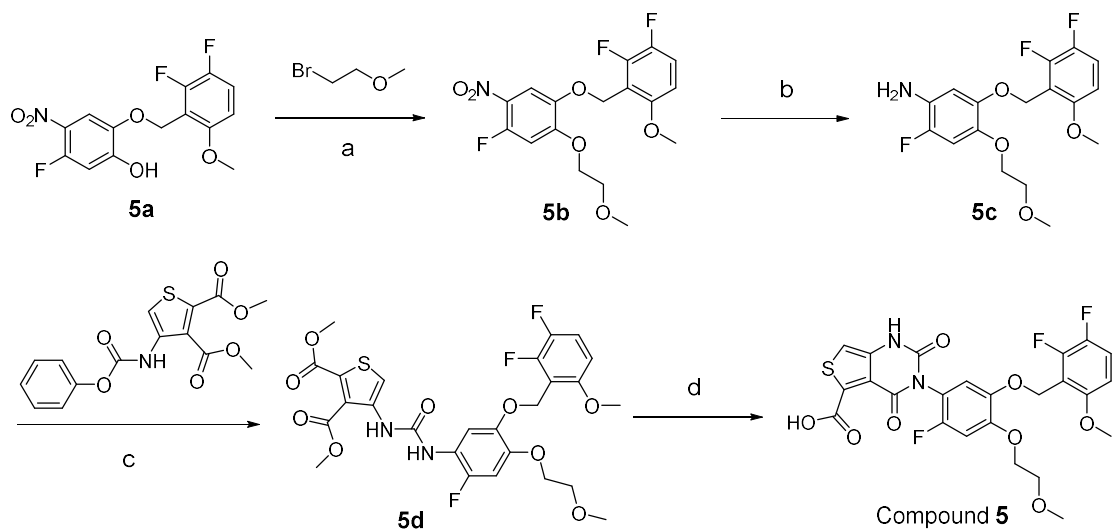

**Fig. 5.** Synthesis route of compound **5**. Reagents and conditions: **a**, KI, K2CO3, ACN, 80 °C, 2 h, sealed tube. **b**, Fe, NH4Cl, EtOH/H2O, 80 °C, 1 h. **c**, Et3N, THF, rt, 16 h. **d**, LiOH, THF/H2O, rt, 1 h.

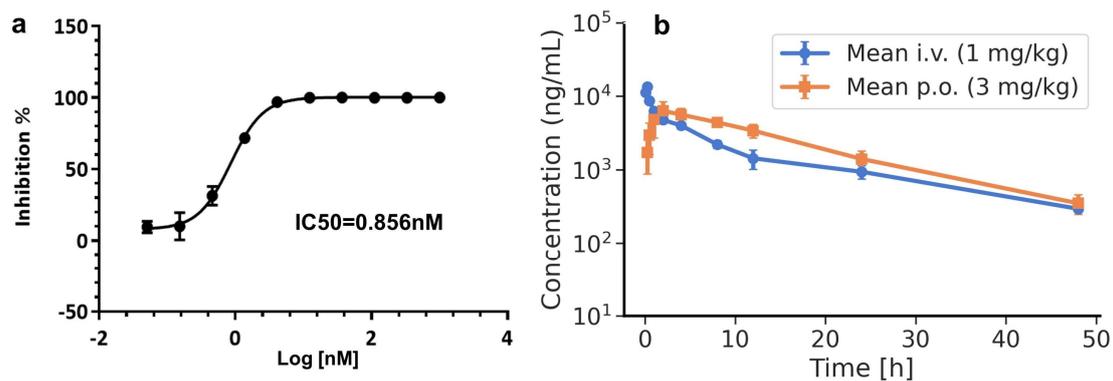

**Fig. 6.** GnRH1R antagonist activity assay and the pharmacokinetic characterization of compound **5**. **a**, Compound **5** antagonist activity against GnRH1R. **b**, Plasma concentrations of compound **5** in rat pharmacokinetic study at doses of 1 and 3 mg/kg for i.v. (blue) and p.o. (orange) treatment, respectively. Measure of center is mean; error bars are s.d.; n = 3 biologically independent animals used for each route of administration.